\documentclass[prl,twocolumn,showpacs]{revtex4}

\usepackage{amsmath}
\usepackage{amsthm}
\usepackage{amssymb}
\usepackage{amscd}
\usepackage[british]{babel}
\usepackage{graphicx}
\usepackage{psfrag}
\usepackage{epsfig}

\def\be{\begin{equation}}
\def\ee{\end{equation}}
\def\bea{\begin{eqnarray}}
\def\eea{\end{eqnarray}}
\def\ba{\begin{array}}
\def\ea{\end{array}}

\begin{document}

\title{F(R) gravity equation of state}

\author{Emilio Elizalde$^{1,}$\footnote{E-mail address: elizalde@ieec.uab.es,
elizalde@math.mit.edu}  and
Pedro J. Silva$^{2,}$\footnote{E-mail address: psilva@ifae.es}}

\affiliation{$^1$Institut de Ci\`encies de l'Espai (CSIC) and Institut
d'Estudis Espacials de Catalunya (IEEC/CSIC) \\
UAB, Ci\`encies, Torre C5-Parell-2a planta E-08193 Bellaterra
(Barcelona) Spain \\
$^2$Institut de Ci\`encies de l'Espai (IEEC-CSIC) and
Institut de F\'{\i}sica d'Altes Energies \\ UAB, E-08193 Bellaterra (Barcelona) Spain}


\thispagestyle{empty}

\begin{abstract}

The ${\bf f}(R)$ gravity field equations are derived as an
equation of state of local space-time thermodynamics. Jacobson's
arguments are non-trivially extended, by means of a more general
definition of local entropy, for which Wald's definition of
dynamic black hole entropy is used, as well as the concept of an
effective Newton constant for graviton exchange, recently appeared
in the literature.

\end{abstract}


\maketitle

\noindent {\it 1.~Introduction.}---In a key article
\cite{Jacobson:1995ab}, Jacobson obtained Einstein's equations (Ee)
from local thermodynamics arguments only. To derive
this result, he generalized black hole (Bh) thermodynamics to
space-time thermodynamics as seen by a local observer. He noted his
finding strongly suggests that, in a fundamental context, Ee are to be
viewed as an equation of state and, therefore, they should
probably not be taken as basic for quantizing gravity. This is consistent
with the idea that gravity is an emergent phenomenon of a more fundamental
framework, like string theory (e.g. \cite{Silva:2005faa}). Were this true,
not only General Relativity, but presumably all generalized gravity
theories should be seen under this same light. 

Modified gravity models constitute a very important dynamical
alternative to $\Lambda$CDM cosmology, in that they have the
capability to describe the current accelerated expansion of our
Universe (dark energy epoch), but also the initial de Sitter phase
and inflation, and even the galaxy rotation curves corresponding
to dark (and ordinary) matter \cite{fr1}. We will here prove that
Jacobson's derivations can be generalized to cover these more
complicated theories of gravity that are extensively used
nowadays. First, we review Jacobson's arguments to introduce the
basic notions and then derive the desired generalization. In
particular, we will completely close the program for
the so called ${\bf f}(R)$ gravities (see e.g.~\cite{fr1}, and
references therein), where the Lagrangian only
depends on the Ricci scalar and its covariant derivatives, leaving
the problem open for more general cases. In \cite{Eling:2006aw},
the field eq.~for ${\bf f}(R)$ of polynomial form were derived
using non-equilibrium thermodynamics arguments (see also \cite{pad1}). Here, we propose
an alternative approach where local thermodynamic equilibrium is
maintained, using the idea of ``local-boost-invariance'' introduced
in \cite{Iyer:1994ys}. 

\noindent {\it 2.~Jacobson's construction in brief.}---Any
free-falling local observer $p$ has some gauge freedom to describe
his local coordinate system. The equivalence principle can be
used to describe space-time in a vicinity of $p$ as flat. Then, we
choose the local space-like area element perpendicular to the
world-line of $p$ to have zero expansion rate $\theta$ and shear
$\sigma$ at a given point on the history of $p$, that we call
$p_0$. In this setting, the past horizon of $p_0$ is called the
``local Rindler horizon'' at $p_0$. Since, locally, we have
Poincar\'{e} symmetry, there is an approximate Killing field $K$
generating boost at $p_0$, vanishing at $p_0$, which we take
future pointing to the inside past of $p_0$.

Having this basic setting, we are ready to give precise meaning to
the local thermodynamic definitions. First, note that local
Rindler horizons are null and act as causal barriers. Therefore,
we can associate entropy $S$ to it, measuring the ``many degrees
of freedom outside'', what presumably results in entanglement
entropy just at the horizon. With this understanding, entropy
is proportional to the area elements of the horizon, where a
fundamental length has to be provided to give an uv cut-off. Heat
$Q$ is energy flow of microscopic degrees of freedom across the
causal barrier, and is felt, therefore, via gravitational energy,
where its source is undetectable. Lastly, the local temperature
$T$ is defined as ``Unruh temperature'', as seen by a local
accelerated observer hovering just inside the horizon. 
Energy flow has to be measured by this same observer, for
consistency.

In more detail, different accelerated observers would measure
different energy flows and temperature, both diverging at the
horizon but with constant ratio, and this is just what will be
used. We have also imposed $\theta=\sigma=0$ at $p_0$, to give a
sort of ``local definition of equilibrium'' since, in general,
causal horizons change in time as they expand and twist. In this
construction, locally and at the instant $p_0$ there is not such a
deformation, and the space is ``at equilibrium". \medskip

\noindent {\it 2.1.~Accelerated observer and approximations.}---On
the above point $p_0$ with its associated local Rindler horizon
$\cal H$, take an accelerated observer hovering just inside the
horizon $\chi$. By the above construction,  $\chi$ is an
approximate local boost Killing field future directed to the past
of $p_0$. Then, the variation of heat (caused by energy flow
across the horizon), measured by $\chi$, is $ \delta Q= \int_{\cal
H}{T_{ab}\chi^b\Sigma^a}$, where the integration is over a pencil
of generators of $\cal H$ at $p_0$. If $K$ is a tangent vector to
the generators of $\cal H$, with affine parameter $\lambda$ such
that $\lambda=0$ at $p_0$, we have that $\chi^a=-k\lambda K^a
+O(\lambda^2)$, where $k$ is the acceleration of $\chi$.
Therefore, $d\Sigma^a=-k\lambda\, d\lambda\, dA$, where $dA$ is
the cross section area element of $\cal H$. Thus, the final
expression for the variation of heat, at leading order, is
\be\label{q}\delta Q= -\int_{\cal H}{k\lambda\,
T_{ab}K^aK^bd\lambda\, dA}\,. \ee Note that for $\chi$ the Unruh
temperature $T$ is set to be \be\label{t}T=k / 2\pi\,.\ee On the
other hand, Jacobson uses that the variation of the entropy
is proportional to the variation of the horizon area $\cal A$,
i.e. $\delta S=\eta \delta {\cal A}$, with $\eta$ an unknown
proportionality constant. Here, $\delta {\cal A}$ measures the
change of the area as we approach the point $p_0$, and therefore
is given by $\delta {\cal A}=\int_{\cal H}{\theta(\lambda)d\lambda
\,dA}$. Next, we use Raychaudhuri`s equation to integrate $\theta$
near $p_0$. In this coordinate system, at
leading order in $\lambda$, we obtain $\theta=-\lambda
R_{ab}K^aK^b+ O(\lambda^2)$, and the relevant expression
for the entropy variation to this order is
\be\label{s}\delta S=-\eta\int_{\cal H}{\lambda R_{ab}K^aK^b
d\lambda\, dA}\,.\ee

\noindent {\it 2.2.~Thermodynamic relations.}---To derive
information of thermodynamic systems, like the equation of state,
we need just the basic thermodynamic relation \be
\label{termo1}\delta Q=T \delta S,\ee and the functional
dependence of $S$ with respect to the energy and size of the
system. In our case we have Eqs.~(\ref{q},\ref{t},\ref{s}) at
disposal, to get the beautiful relation \be
\label{Eekk}T_{ab}K^aK^b={1\over 2\pi}\eta R_{ab}K^aK^b.\ee Since
$K$ is an arbitrary null vector on $\cal H$, we can write the
unprojected equation $ T_{ab}={1\over 2\pi}\eta R_{ab}+g_{ab}h$,
with $h$ an unknown function, arbitrary as of now. Using then that
the lhs is divergence-free, plus the Bianchi identities for the
Ricci tensor, we get the integrability conditions: ${1\over
4\pi}\eta \nabla_aR=-\nabla_a h$, and therefore the final form of
the thermodynamic relation is \be
\left({2\pi\over\eta}\right)T_{ab}=\left( R_{ab}-{R\over
2}g_{ab}\right) +\Lambda g_{ab}, \ee where $\Lambda$ is an
integration constant.

To summarize, we have here obtained the Ee as an equation of state
for a local free-falling observer. To deduce the above, we have
used the following critical assumptions: (i) Measurements are done
in a vicinity of a general point $p_0$. (ii) Our local coordinate
system is at equilibrium, in the sense that $\theta,\sigma=0$ at
$p_0$. (iii) The accelerated observer $\chi$ tends to $K$, a null
vector generator of the causal horizon. (iv) We always restrict
ourselves to the leading order approximation in the affine
parameter $\lambda$. \vspace{0.5mm}

\noindent {\it 3.~The general case of modified gravity.}---We 
apply the above construction to more general theories of
gravity. Following Iyer and Wald \cite{Iyer:1994ys}, we just
assume that our Lagrangian is diffeomorphism invariant, in a
$n$-dimensional oriented manifold $\cal M$,  being the dynamical
fields a Lorentz signature metric $g_{ab}$ and other matter
fields $\psi$. The most general Lagrangian is 
 \bea {\bf L}&=&{\bf L}\left(g_{ab},R_{cdef},\nabla_{a_1}
R_{cdef},\ldots,\nabla_{(a_1}\ldots\nabla_{a_n)}R_{cdef}, \right.
\nonumber \\ && \left.
\psi,\nabla_{a_1}\psi,\ldots,\nabla_{(a_1}\ldots\nabla_{a_n)}\psi\right).
\eea The corresponding field equations con be found by a
variational procedure on $(g_{ab},\psi)$, so that we get \be
\delta {\bf L} = \epsilon\left({\bf E}_g^{\;ab}\delta g_{ab}+{\bf
E}_\psi \delta\psi\right) +d{\bf \Theta}, \ee where $\epsilon$ is
the volume element and ${\bf \Theta}$ a $(n-1)$-form. Hence, the
field equations of the theory are: ${\bf E}_g^{\;ab}=0,\ {\bf
E}_\psi=0$. In \cite{Iyer:1994ys}, it was found how to write them
from a variation of the $(g_{ab},R_{cdef})$, as if they were
independent variables, so that we get, after the corresponding
identifications, \be {\bf
E}_g^{\,ab}=A_g^{\,ab}+E_R^{\;pqra}R_{pqr}^{\quad
b}+2\nabla_p\nabla_q E_R^{\;pabq}, \ee where
$(A_g^{\;ab},E_R^{\;pabq})$ are the variations of $\bf L$ with
respect to $(g_{ab},R_{pabq})$ in each case, taken as independent
variables. In the above expressions, if the derivatives of
$R_{cdef}$ occur in the Lagrangian, one integrates by parts and
then takes its variation, to obtain $E_R^{\;pabq}$.

This form of the field equations is useful due to its relation to
Bh thermodynamics. Basically, it has been known for a while now
\cite{Wald:1993nt}, that in the case when we have a stationary Bh
solution, the entropy $S$ can be calculated as a Noether charge
evaluated at the bifurcation $(n-2)$-surface of the event-horizon
$\Sigma$. In these cases, the entropy is given by \be \label{nc}
S=-2\pi\int_\Sigma E_R^{\;abpq}\epsilon_{ab}\epsilon_{pq},\ee
where $\epsilon_{ab}$ is the binormal vector of $\Sigma$.

What is less understood is the case of dynamical Bh entropy.
There, the event-horizon is not bifurcated and, therefore, the
above formula does not hold. Nevertheless, in \cite{Iyer:1994ys} a
prescription that passes the basic tests of consistency for the
corresponding entropy is presented, although it is not completely
clear if that is a good answer. In any case, the idea is to
approximate the metric $g$, in a vicinity of a given point $p$ of
the event-horizon, by a boost-invariant metric $g^{Iq}$. This is
done by altering the original Taylor expansion of the metric
around $p$, so that the new metric is boost invariant up to some
order $q$, that defines the size of the vicinity where our
approximation is valid. Then, for this boost invariant metric,
there is a Killing vector field that, on the horizon, is null, and
vanishes at $p$. We thus have created an approximated
bifurcation surface of order $q$ and can use the same
expression as before for the entropy, only that the integration is
done on the boost-invariant variables:
\be
S_{dyn}(\Sigma_p)=-2\pi\int_{\Sigma_p} \hat
E_R^{\;abpq}\epsilon_{ab}\epsilon_{pq},\ee where $\hat
E_R^{\;abpq}=E_R^{\;abpq}(g^{I_q})$ (see \cite{Iyer:1994ys} for details).

Having understood these modifications for calculating the
Bh entropy, we are almost ready to continue. Still, some
information on the geometrical meaning of Eq.~(\ref{nc}).
In \cite{Brustein:2007jj} it was noticed that, for a
static Bh, entropy can always be re-expressed as the area of
the bifurcation $(n-2)$-surface $A$ divided by 4 in units of an
effective Newton constant $G_{eff}$, i.e. \be
\label{Geff}\hspace*{-3mm} S={A\over
4G_{eff}}\quad\hbox{where}\quad {1\over 8\pi
G_{eff}}=E_R^{\;abpq}\epsilon_{ab}\epsilon_{pq}.\ee The above
result has been checked for some string theory cases where it was
found that $G_{eff}$ is indeed constant on the bifurcation
surface. This result has to be supplemented with the key
observation that the above effective Newton constant plays also
the role of an effective gravitational coupling for graviton
exchange. In other words, the kinetic term of the $n$-dimensional
graviton, obtained from the general Lagrangian $\bf L$, is
precisely of the form ${1\over 4}E_R^{\;abpq}\left(\nabla_r h_{bq}
\nabla^r h_{ap}+\ldots\right)$ and, hence, $E_R^{\;abpq}$ can be
thought of as the strength of the graviton interaction in all
possible polarizations. In retrospective, $G_{eff}$ corresponds to
the strength of the gravitational interaction in the particular
polarizations defined by the binormal of the bifurcation
$(n-2)$-surface $A$.\vspace{0.5mm}

\noindent {\it 3.1.~Field equation as equation of state.}---Now
that we have the basic inputs for the possible geometrical
interpretation of the Bh entropy $S$ for generalized theories of
gravity, we are ready to consider the problem of defining there 
a local version of Bh thermodynamics, following the steps
of Jacobson. Note that all definitions regarding local
observers (accelerated or not), local Rindler horizons, and so on,
are based on differential geometry and the equivalence principle.
We expect all these definitions to hold in the general
setting and, thus, we leave them unchanged.

What makes the difference, being one of the key points in this
generalization, is the definition of the local entropy. After the
discussion above, it seems natural to relate
entropy to the area of the causal horizon, only that now we
replace the proportionality constant with a field-dependent
effective constant. In other words, we state that the local
variation of the entropy is still proportional to the variation of
the area of the causal horizon, but in units of this effective
Newton constant. Therefore, we write now \footnote{We thank T.
Jacobson for pointing out that in a first version of this work,
$\delta S$ assumed $\eta_e$ constant along $K$ before using the
boost-invariant truncation.}
\be \delta S=\delta\,(\eta_{e} A),\ee
where $\eta_e$ is, in general, a function
of the metric and its derivatives to a given order $l+2$, i.e.
\be\eta_e=\eta_e\left(g_{ab},R_{cdef},\nabla^{(l)}
R_{pqrs}\right)\,.\ee
Using the above {\it ansatz}, we are ready to proceed with our
derivation. Since we just change the definition of entropy
variation, due to the energy flow across the local Rindler
horizon, we get the modified expression
\be\label{s17}\delta S=-\int_{\cal H}{\lambda\,(\eta_e \,R_{ab}
-\nabla_{a}\nabla_{b}\eta_e)K^aK^b d\lambda\, dA}+
O(\lambda^2)\,.\ee
It is important to notice that, in this expression,
$(\eta_e,k^a\nabla_a\eta_e)$ are to be evaluated at its {\it
leading} contribution in $\lambda$. We have used its
boost-invariant part at first order in lambda to effectively
incorporate the boost invariant notion of \cite{Iyer:1994ys}
creating an ``approximated bifurcation point at first order in
$\lambda$'' at $p_0$.

The other part of the derivation is unaffected and gives the same
result of (\ref{q}), namely \be\delta Q= -\int_{\cal H}{k\lambda\,
T_{ab}K^aK^bd\lambda\, dA} + O(\lambda^2)\,. \ee Therefore, the
thermodynamic relation (\ref{termo1}) implies
\be T_{ab}K^aK^b={1\over 2\pi}(\eta_e
R_{ab}-\nabla_{(a}\nabla_{b)}\eta_e)K^aK^b\,.\ee
At this point we consider the general differential equation,
removing the contraction with $K$, thus
\be T^{ab}={\eta_e\over 2\pi}
R^{ab}-\nabla^{(a}\nabla^{b)}{\eta_e\over 2\pi}+g^{ab}H\,,\ee
where the new terms are added based on the fact that $K$ is a
tangent vector of the null geodetics at $p_0$, generating local
boost. Hence, at this point we have a local equation with two
unknown functions of the metric and its derivatives $(\eta_e,H)$.

To find the form of these three functions, we use the
integrability condition
\be \label{divE}\nabla_a\left({\eta_e\over 2\pi}
R^{ab}-\nabla^{(a}\nabla^{b)}{\eta_e\over 2\pi}+g^{ab}H\right)=0,
\ee
obtained from the observation that the rhs should be divergence
free. After some algebra,
\bea 0&=&{\eta_e\over
4\pi}\nabla^bR+\left(\nabla_a\nabla^b\nabla^a-\nabla^b\nabla^2\right){\eta_e\over
2\pi}+\nabla^bH \nonumber \\ &&-
{1\over2}\left(\nabla^2\nabla^b+\nabla_a\nabla^b\nabla^a
\right){\eta_e\over 2\pi}. \eea At this point, with no lose of
generality, we can set \be H=h+\nabla^2{\eta_e\over 2\pi}, \ee
finally obtaining the expression \bea  0={\eta_e\over
4\pi}\nabla^bR+\nabla^bh . \eea

\noindent {\it 3.2.~The specific case of f(R)
gravities.}---Eq.~(\ref{divE}) can in principle be solved in many
different ways. Here we will consider the simplest possibility,
that eventually leads to the so-called ${\bf f}(R)$ gravities
\cite{fr1}, a special---but phenomenologically very
important---family of the general theories of modified gravity
where only the Ricci scalar is involved, as well as their
covariant derivatives, i.e. $ {\bf L}={\bf f}(R,\nabla^n R)$.

The solution of Eq.~(\ref{divE}) we are considering is perhaps the
simplest one where the first two terms cancel each other. The
last one can be easily integrated assuming $\eta_e$
is a function of $R$ only, thus $ h=-{1\over2}{\bf
f}(R)$, with ${\eta_e\over 2\pi}={\partial {\bf f}\over \partial
R}$. Therefore, the final form once we have
collected all terms in the above derivation is \be \label{fE1}
E^{ab}={\partial {\bf f}\over
\partial R}R^{ab}-\nabla^{(a}\nabla^{b)}{\partial {\bf f}\over
\partial R} + g^{ab}\left(-{1\over2}{\bf f}+\nabla^2{\partial {\bf
f}\over
\partial R}\right), \ee ${\bf f}$ a function of $R$ and its
covariant derivatives only.

Eq.~(\ref{fE1}) is in fact the correct field theory equation for
${\bf f}(R)$ gravities, provided we identify the function {\bf
f}(R) as the Lagrangian of the theory. Also, in this case the
effective Newton constant of (\ref{Geff}) is related to $\eta_e$,
as is expected from the relation
\bea {1\over 8\pi G_{eff}} &=&
E_R^{pqrs}\epsilon_{pq}\epsilon_{rs} ={\partial {\bf
f}\over\partial R}(g^{pr}g^{qs}-g^{qr}g^{ps})\epsilon_{pq}\epsilon_{rs} \nonumber \\
&=&{\partial {\bf f}\over \partial R}={\eta_e\over 2\pi} \eea
Note that, for these theories, the different polarizations of the
gravitons only enter in the definition of the effective Newton
constant through the metric itself. This is an important
simplification that, in turn, permits to find the solution of the
integrability condition (\ref{divE}). In retrospective, and to
summarize, we have succeeded in our thermodynamic derivation of
${\bf f}(R)$ gravities where, remarkably, exactly as in the case
of Einstein gravity, {\it the local field equations can be thought
of as an equation of state}.

It will be very interesting to see if this derivation can be
extended to the more complicated cases, stemming from string
theory, where the full Riemann tensor is involved in the
Lagrangian. This seems to imply a sort of tetrad decomposition of
the effective Newton constant such that one recovers, at the end,
only the polarization normal to the causal barrier of the local
Rindler horizon. Work along this line is in progress. As a last
comment, in our derivation we have used the first law, but no
information is given bout the second law. In fact it is not know
if the second law is present in generalized gravities (see
\cite{Jacobson:1995uq}). \vspace{0.5mm}

We thank S.~Odintsov and E.~Verlinde for useful discussions. Work
partially funded by Ministerio de Educaci\'on y Ciencia, Spain,
projects CICYT-FEDER-FPA2005-02211 and FIS2006-02842, CSIC under
the I3P program, and AGAUR,
contract 2005SGR-00790.


\end{document}